\documentclass{article}
\usepackage[utf8]{inputenc}
\usepackage{authblk}
\usepackage{hyperref}
\usepackage{bm}
\usepackage{amsmath}
\usepackage{amsfonts}
\usepackage{amssymb}
\usepackage{array}

\title {Lorentz, Poincaré, and Einstein: Rethinking Doppler, Aberration, and the Fresnel Drag}
\author{Galina Weinstein\thanks{The Department of Philosophy, University of Haifa.}}

\begin{document}

\maketitle

\begin{abstract}
This paper examines Lorentz’s 1895 derivations of the classical Doppler formula and Fresnel drag, Einstein’s 1905 derivation of the relativistic Doppler effect and aberration, and Einstein’s 1907 kinematical route to the exact velocity composition law from which Fresnel drag is obtained as a low-velocity limit. 
Einstein acknowledged that he had read Lorentz’s \emph{Versuch} well before 1905. In 1907, Einstein identified Lorentz’s \emph{Versuch} as a crucial precursor to relativity. In that work, Lorentz had already invoked local time to derive Fresnel’s drag coefficient from Maxwell’s equations.
There is a genuine “family resemblance” between Lorentz’s and Einstein’s treatments in that both preserve the phase of a plane wave under transformation. Yet I demonstrate that this resemblance is only formal. 
I also discuss the absence of the relativistic Doppler and aberration laws in Poincaré’s Dynamics of the Electron.
\end{abstract}

\section{Introduction}

This paper examines Lorentz’s 1895 derivations of the classical Doppler formula and Fresnel drag, Einstein’s 1905 treatment of the relativistic Doppler effect and stellar aberration, and his 1907 kinematical derivation of the exact velocity composition law from which Fresnel drag emerges as the low-velocity limit. Einstein had read Lorentz’s \emph{Versuch} well before 1905 \cite{CPAE1}, Doc. 130; by 1902, he had studied the treatise in which Lorentz introduced local time to recover Fresnel’s coefficient from Maxwell’s equations. In his 1907 review, "On the Relativity Principle and the Conclusions Drawn from It," Einstein identified Lorentz’s 1895 \emph{Versuch} as a crucial precursor to relativity, noting that although it still assumed an immobile ether, it won empirical support by correctly yielding Fresnel’s drag coefficient in agreement with Fizeau’s experiment \cite{Einstein07}.

By following the Doppler effect (longitudinal and transverse), aberration, Fresnel drag, and the velocity addition law side by side, I show that the apparent algebraic continuities conceal a decisive conceptual break. Einstein does not merely repair Lorentz’s formulas; he replaces the premises from which those formulas were drawn, and in so doing parts ways with both Lorentz and Poincaré.  

The history of the Doppler effect has been thoroughly examined in \cite{Dar} and will not be pursued further here; see also \cite{Nolte} for its early reception, later vindication, and modern applications. For the histories of the Fizeau experiment, Fresnel dragging, and aberration, see \cite{JS}.

\section{Doppler and Aberration}

\subsection{Asymmetric Classical Doppler Effect in Ether Theory} \label{Asym}

The classical Doppler effect in an ether theory involves two factors: 
(1) how the source emits wave crests into the ether, and 
(2) how the observer encounters them while moving through the ether. 
This gives two basic formulas, one for approaching motion and one for receding motion:

\begin{equation} \label{Dop1}
\nu' \;= \; \nu \left( \frac{c + u}{c - v}\right), \qquad \text{Approaching motion}, 
\end{equation}

\begin{equation} \label{Dop2}
\nu' \;= \; \nu \left( \frac{c - u}{c + v}\right), \qquad \text{Receding motion},
\end{equation}
where $c$ is the wave speed in the ether, $u$ the observer’s velocity through the ether, $v$ the source velocity through the ether, $\nu$ the emitted frequency, and $\nu'$ the observed frequency. 

\medskip

\noindent\textit{First case: moving source, stationary observer ($u=0$).}  
The source emits crests into the ether as it moves at speed $v$. If the source recedes, the crests are spaced farther apart, and the frequency decreases; if it approaches, the crests bunch together, and the frequency increases. For sound, this corresponds to a change in pitch; for light (if carried by an ether), to a color shift (redshift or blueshift).  
From equation \eqref{Dop1}, the approaching case is:
\begin{equation} \label{ETD}
\nu' \;= \; \nu \left(\frac{c}{c - v} \right).    
\end{equation}
As $v \to c$, the denominator vanishes and $\nu' \to \infty$. For sound waves, this is the precursor to a shock. Once the source surpasses the speed of sound, the familiar picture of wave crests bunching up in front of the source breaks down. For light, this leads to an unphysical singularity of infinite frequency and vanishing wavelength.

\noindent From equation \eqref{Dop2}, the receding case is:
\begin{equation} \label{ETD-1}
\nu' \;= \; \nu \left(\frac{c}{c + v} \right).    
\end{equation}
As $v \to c$, $\nu' \to \nu/2$, the observed pitch halves (or wavelength doubles).  
\medskip

\noindent\textit{Second case: stationary source ($v=0$), moving observer.}  
The source emits crests at a steady rate $\nu$. A moving observer encounters them more frequently as they approach and less frequently as they recede.  

\noindent From equation \eqref{Dop1}, for approach ($u>0$):
\begin{equation} \label{2n}
\nu' \;= \; \nu \left(1 + \frac{u}{c} \right).    
\end{equation}
At $u \to c$, $\nu' = 2\nu$, the observed pitch doubles.  

\noindent From equation \eqref{Dop2}, for recession:
\begin{equation} \label{0n}
\nu' \;= \; \nu \left(1 - \frac{u}{c} \right).    
\end{equation}
At $u \to c$, $\nu' \to 0$, no crests are encountered (silence or darkness).  

\medskip

\noindent In summary, motion of the \emph{observer} produces a bounded effect (frequency between $0$ and $2\nu$), while motion of the \emph{source} leads to extreme predictions, including a singularity as $v \to c$. This asymmetry arises because the ether selects a preferred rest frame, distinguishing source motion from observer motion. 

At first glance, the classical Doppler formulas \eqref{Dop1} and \eqref{Dop2} look nicely symmetric because the observer’s speed $u$ appears in the numerator, and the source’s speed $v$ appears in the denominator. This seems to treat the two roles, observer and source, on the same footing, just as in Einstein’s relativity.  
But the resemblance is misleading. In the classical (ether) picture, the two terms mean very different things. The $u$-term reflects how the observer moves through the medium and intercepts the crests, while the $v$-term reflects how the source lays down the crests into the medium. The symmetry is only mathematical; it depends on an absolute frame, the ether, in which the wave speed $c$ is defined.  

Einstein removed the ether and the asymmetry. In relativity, only relative motion matters, so the Doppler effect has the same formula whether the source or observer moves. 
In his longitudinal Doppler law \eqref{DopL}, there is no need for a separate observer speed $u$ and source speed $v$. Only the relative velocity between them matters. Thus, Einstein’s formula involves just $v$ (the relative speed) and $c$, not both $u$ and $v$. What looks like the same symmetry in the classical equations is in fact a disguised asymmetry, because it rests on the hidden assumption of the ether. 
The relativistic law \eqref{DopL} still predicts ever-larger shifts as relative speed approaches $c$, but the divergence at $v=c$ is unreachable, since no material object can attain the speed of light (see the discussion in section \ref{EDR})

\subsection{Lorentz Derives the Classical Doppler Effect} \label{CD}

In his 1895 treatise, Lorentz begins with the plane wave of phase \eqref{Lor-Phi} written in the ether rest frame. In other words, the source is taken as stationary in the ether and the observer is moving through the ether with velocity $-\mathfrak{p}$. This is the second case discussed in the previous section \ref{Asym}.

He introduces moving coordinates \eqref{sigma}, and substitutes them into the phase.%
\footnote{I shall not discuss here the work of Woldemar Voigt and its possible relation to Lorentz’s later formulations. Readers interested in this subject are referred to Abraham Pais’s account in \emph{Subtle is the Lord} (see Chapter 6), where he reviews Voigt’s 1887 transformations, Lorentz’s delayed awareness of them, and Minkowski’s later comments \cite{Pais}.}
This yields equation \eqref{phicosT}, where $\mathfrak{p_n}$ is the projection of the $(\Sigma,Y,Z)$-frame velocity along the wave normal \eqref{normal}; the observer’s velocity is $-\mathfrak{p_n}$. 
At fixed $(\Sigma,Y,Z)$, the oscillation period $T'$ is defined by the condition that the phase advances by $2\pi$ between $t$ and $t+T'$. Since the coefficient of $t$ is $1 + \mathfrak{p}_n/V$ [see equation \eqref{fixed}], the observed period is $T'$ \eqref{Tfra}, hence the observed frequency is $\nu'$ \eqref{nuu}. Equation \eqref{FT} is Lorentz’s form of the classical Doppler principle (to first order in $\mathfrak{p}/V$).

Lorentz begins by expressing a plane wave in an immobile ether \cite{Lorentz1895}:

\begin{equation} \label{Lor-Phi}
\Phi \;=\; A \cos \frac{2\pi}{T}
\left( t \;-\; \frac{a_x x + a_y y + a_z z}{V} \;+\; p \right).
\end{equation}
Here, $T$ is the oscillation period, $V$ the propagation speed of light in the ether,  $a_x, a_y, a_z$ the direction cosines of the wave normal, and $p$ a constant phase.

Now he introduces new coordinates  $\Sigma, Y, Z$ adapted to an observer moving at velocity $-\mathfrak{p}$ relative to the ether \cite{Lorentz1895}:

\begin{equation} \label{sigma}
x = \Sigma - \mathfrak{p}_x t, 
\qquad y = Y - \mathfrak{p}_y t, 
\qquad z = Z - \mathfrak{p}_z t .
\end{equation}
These are Galilean transformations between the ether frame $(x, y, z, t)$ and the moving observer’s frame $(\Sigma, Y, Z, t')$, where $t=t'$.  

\noindent Plugging the coordinate transformation \eqref{sigma} into the phase \eqref{Lor-Phi} yields \cite{Lorentz1895}:%
\footnote{We substitute the coordinate transformation \eqref{sigma} into the linear form in the numerator of the phase \eqref{Lor-Phi}:
\begin{equation}
a_x x + a_y y + a_z z
= a_x(\Sigma - \mathfrak{p}_x t) + a_y(Y - \mathfrak{p}_y t) + a_z(Z - \mathfrak{p}_z t).
\end{equation}
Then, we expand:

\begin{equation}
a_x x + a_y y + a_z z
= (a_x \Sigma + a_y Y + a_z Z) - (a_x \mathfrak{p}_x + a_y \mathfrak{p}_y + a_z \mathfrak{p}_z)\, t .
\end{equation}
We introduce the projection of $\mathfrak{p}$ along the wave normal \eqref{normal}. Thus: 

\begin{equation} \label{xyzpn}
a_x x + a_y y + a_z z
= (a_x \Sigma + a_y Y + a_z Z) - \mathfrak{p}_n t .
\end{equation}
Now we plug back the linear form \eqref{xyzpn} into the phase \eqref{Lor-Phi}, and the phase becomes
\begin{equation}
\Phi \;=\; A \cos \frac{2\pi}{T}
\left(t - \frac{(a_x \Sigma + a_y Y + a_z Z) - \mathfrak{p}_n t}{V} + p \right).
\end{equation}
We then separate terms:

\begin{equation}
\Phi \;=\; A \cos \frac{2\pi}{T}
\left(t - \frac{a_x \Sigma + a_y Y + a_z Z}{V}
+ \frac{\mathfrak{p}_n}{V} t + p \right).
\end{equation}
This finally yields equation \eqref{phicosT}.}

\begin{equation} \label{phicosT}
\Phi \;=\; A \cos \frac{2\pi}{T}
\left( t + \frac{\mathfrak{p}_n}{V} t \;-\; \frac{a_x \Sigma + a_y Y + a_z Z}{V} \;+\; p \right), \qquad \text{where:}
\end{equation}

\begin{equation} \label{normal}
\mathfrak{p}_n \;=\; a_x \mathfrak{p}_x + a_y \mathfrak{p}_y + a_z \mathfrak{p}_z.
\end{equation}
is the component of the velocity along the wave normal.

\noindent Equation \eqref{phicosT} describes the wave phase expressed in the moving observer’s coordinates $(\Sigma, Y, Z)$. The first two temporal terms:

\begin{equation}
t + \frac{\mathfrak{p}_n}{V} t,    
\end{equation}
show that the frequency is shifted compared to the ether frame, and the spatial part:

\begin{equation}
\frac{a_x \Sigma + a_y Y + a_z Z}{V},    
\end{equation}
represents the propagation of the wavefronts in the observer’s coordinate system.

We want to know what an observer at rest in the moving frame $(\Sigma, Y, Z)$ measures. What are the oscillations of the wave measured by this particular observer? At the moving coordinates $(\Sigma, Y, Z)$, the observer is at rest. That means their coordinates do not change with time $\Sigma, Y, Z = $ const. So the only variable that changes for them is the time $t$. Then the phase \eqref{phicosT} reduces to purely a function of $t$ only: 

\begin{equation} \label{fixed}
\Phi \;=\; A \cos \frac{2\pi}{T}
\left( (1+ \frac{\mathfrak{p}_n}{V})t \;+\; \text{const.} \right).
\end{equation}
Lorentz asks after what interval of time $T'$ does the phase advance by $2 \pi$ for the moving observer?
The wave repeats itself whenever the argument of the cosine increases by $2\pi$.
So let $T'$ be the observed oscillation period. Then we must have:
\begin{equation}
\frac{2\pi}{T}\left(1 + \frac{\mathfrak{p}_n}{V}\right) T' = 2\pi .
\end{equation}
Canceling $2\pi$ gives:
\begin{equation}
\frac{1}{T}\left(1 + \frac{\mathfrak{p}_n}{V}\right) T' = 1 .
\end{equation}
Solving for $T'$ yields the observed oscillation period as measured by the moving observer who is at rest in the $(\Sigma, Y, Z)$ system \cite{Lorentz1895}:
\begin{equation} \label{Tfra}
T' \;=\; \frac{T}{1 + \tfrac{\mathfrak{p}_n}{V}} .
\end{equation}
where $T$ is the “true” oscillation period of the wave in the ether frame and $T'$ is the "apparent" period for a fixed observer in the moving frame.
Equivalently, for the frequency $\nu' = 1/T'$:
\begin{equation} \label{nuu}
\nu' \;=\; \nu \left( 1 + \frac{\mathfrak{p}_n}{V} \right).
\end{equation}
This is Lorentz’s classical Doppler expression [equivalent to equations \eqref{2n} and \eqref{0n}].

\noindent Doing a first–order Taylor expansion in the small parameter $\dfrac{\mathfrak{p}}{V}$, assuming $\mathfrak{p}_n \ll V$, yields \cite{Lorentz1895}:%
\footnote{Notice that the denominator of $T'$ \eqref{Tfra} makes it nonlinear in $\frac{\mathfrak{p}_n}{V}$ because expanding it contains all powers of $\frac{\mathfrak{p}_n}{V}$. However, the first-order approximation keeps only the linear term in $\frac{\mathfrak{p}_n}{V}$ [in \eqref{T'}]. On the other hand, the exact expression for $\nu'$ \eqref{nuu} is already linear in $\frac{\mathfrak{p}_n}{V}$, because no higher-order terms appear when expanding it. Thus, the approximation \eqref{FT} does not change it.}

\begin{equation} \label{T'}
T' \;\simeq\; T \left( 1 - \frac{\mathfrak{p}_n}{V} \right), 
\end{equation}
\begin{equation} \label{FT}
\nu' \;=\; \nu \left( 1 + \frac{\mathfrak{p}_n}{V} \right).
\end{equation}
With $\mathbf{a} = (a_x, a_y, a_z)$ the direction cosines of the wave normal (in the direction of propagation), the sign of $\mathfrak{p}$ determines the Doppler shift. 

If $\mathfrak{p}_n > 0$, i.e.,\ the observer is moving toward the source along the wave normal, then $\nu' > \nu$, so the observed frequency is increased (blue-shifted). Conversely, if $\mathfrak{p}_n < 0$, i.e.\ the observer is receding along the wave normal, then $\nu' < \nu$, so the observed frequency is decreased (red-shifted).
If $\nu' = 0$, then this requires:
\begin{equation}
1 + \frac{\mathfrak{p}_n}{V} = 0 
\;\;\;\;\Longrightarrow\;\;\;\; 
\mathfrak{p}_n = -V .
\end{equation}
As explained in section \ref{Asym}, this corresponds to the observer moving away from the source precisely at the wave speed. 
Physically, the wave crests never catch up to the observer, and the observed frequency 
drops to zero---the oscillations disappear. This is therefore the maximum possible redshift 
in the classical Doppler formula (since negative frequencies are not physical here). This is paradoxical because an observer receding at the speed of the wave sees no oscillations at all. However, light cannot just “disappear” by changing frames. The darkness paradox highlights the internal inconsistency of the ether-based view, and is discussed in detail in section \ref{Dark}.

What Lorentz derived is the \emph{classical Doppler effect} for a moving observer in a stationary medium (ether), to first order in $\dfrac{\mathfrak{p}}{V}$. So, for the opposite case, $\mathfrak{p}_n = +V$, the formula \eqref{FT} gives $\nu' = 2\nu$ (see explanation in section \ref{Asym}). That would also be problematic, as it represents the maximum blueshift in this simple linear theory.  

\subsection{Einstein Derives the Relativistic Doppler and Aberration Formulas} \label{EDR}

In section \S7 of his 1905 relativity paper, Einstein begins in the “rest” frame $K$ with the electromagnetic plane wave of phase $\Phi$ \eqref{Ph}. Substituting the inverse Lorentz transformation \eqref{TI} into $\Phi$ yields the transformed phase \eqref{Phi}. By the principle of relativity, the phase in $k$ must retain the form \eqref{Ph-k}. Equating coefficients of $\tau, \xi, \eta, \zeta$ in \eqref{Ph-k} with those in \eqref{Phi}, Einstein obtains from the $\tau$ term the frequency relation \eqref{w}, and from the spatial terms \eqref{eqa}, \eqref{b'}, \eqref{c'}. With $l=\cos\varphi$, these give the Doppler and aberration laws \eqref{Dop} and \eqref{Ab}: 

\begin{enumerate}
    \item Einstein starts with a plane electromagnetic wave in the system $K$%
\footnote{Written as:

\begin{equation} \label{Ph-1}
X = X_0 \sin\Phi,\quad
Y = Y_0 \sin\Phi,\quad
Z = Z_0 \sin\Phi,    
\end{equation}

\begin{equation} \label{Ph-2}
L = L_0 \sin\Phi,\quad
M = M_0 \sin\Phi,\quad
N = N_0 \sin\Phi,    
\end{equation}.}
with phase \cite{Einstein05}:
\begin{equation} \label{Ph}
\Phi \;=\; \omega\!\left(t - \frac{l x + m y + n z}{c}\right). 
\end{equation}
Here, $c$ is the velocity of light, $l,m,n$ are the direction cosines of the wave normal, and $\omega=2 \pi \nu$ is the angular frequency.

\item He asks how these waves appear in the moving system $k$?  
He applies the Lorentz coordinate transformations from section \S3 (for $x,y,z,t$):

\begin{equation} \label{eq12}
\tau = \gamma\!\left( t - \frac{v}{c^2}\,x \right), \quad
\xi = \gamma\,(x - v t), \quad
\eta = y, \quad
\zeta = z, \quad \gamma=\frac{1}{\sqrt{1-\frac{v^2}{c^2}}}
\end{equation}
He inverts the transformations \eqref{eq12} to express $t, x$ by $\tau, \xi$ to get the inverse Lorentz transformation:
\begin{equation} \label{TI}
t \;=\; \gamma\!\left(\tau + \tfrac{v}{c^{2}}\,\xi\right), 
\qquad 
x \;=\; \gamma\!\left(\xi + v \tau\right), \qquad y \;=\; \eta, \qquad z \;=\; \zeta.
\end{equation}
and then substitutes the inverse Lorentz transformation \eqref{TI} into the phase \eqref{Ph}: 
\begin{equation} \label{Phi}
\begin{aligned}
\Phi
  &= \omega \Bigg[
     \gamma\!\left(\tau + \frac{v}{c^{2}}\xi\right)
     - \frac{l\,\gamma(\xi + v\tau)}{c}
     - \frac{m}{c}\eta
     - \frac{n}{c}\zeta
     \Bigg] \\[6pt]
  &= \omega \Bigg[
     \underbrace{\gamma\!\left(1 - \tfrac{l v}{c}\right)}_{\text{coeff of }\tau} \,\tau
     + \underbrace{\gamma\!\left(\tfrac{v}{c^{2}} - \tfrac{l}{c}\right)}_{\text{coeff of }\xi} \,\xi
     - \tfrac{m}{c}\,\eta
     - \tfrac{n}{c}\,\zeta
     \Bigg].
\end{aligned}  
\end{equation}

Einstein considers two inertial systems: 

I) \textit{system $K$} is the "rest" system in which the source is described and the plane electromagnetic wave is written with phase \eqref{Ph}. 

II) \textit{system $k$} moves with constant velocity $v$ in the $+x$--direction relative to $K$. 

Coordinates in $K$ are $(x,y,z,t)$ and in $k$ are $(\xi,\eta,\zeta,\tau)$. 

The Lorentz transformation \eqref{eq12} carries events from $K$ to $k$; we use its inverse \eqref{TI} to express $(x,t)$ by $(\xi,\tau)$ and substitute into the phase \eqref{Ph}. This gives the transformed phase \eqref{Phi} written in $k$-coordinates, but still with the $K$-parameters $(\omega, l, m, n)$. 

\item \label{same} According to the principle of relativity, equation \eqref{Ph} in system $K$ can be written in the plane-wave form in system $k$:

\begin{equation} \label{Ph-k}
\Phi \;=\; \omega'\!\left(\tau - \frac{l' \xi + m' \eta + n' \zeta}{c}\right)  \;=\; \omega' \tau-\frac{\omega'}{c}\!\left(l' \xi + m' \eta + n' \zeta\right).   
\end{equation}

By the relativity principle, a plane wave must again retain the same form in any inertial system. Hence, in $k$ the same physical wave has the canonical form \eqref{Ph-k} with its own frequency $\omega'$ and its own direction cosines $(l', m', n')$ of the wave normal measured in $k$. 

\item Now he equates like terms, the coefficients of $\tau, \xi, \eta, \zeta$. Terms in the equation \eqref{Phi} derived by the inverse Lorentz transformation \eqref{TI} $=$ terms in equation \eqref{Ph-k}.
We start from the coefficient of $\tau$ \cite{Einstein05}:

\begin{equation} \label{w}
\boxed{\omega' = \omega \gamma\!\left(1 - \frac{l v}{c}\right).}  
\end{equation}
Then we calculate the coefficient of $\xi$:%
\footnote{\label{Doppler} We calculate:
\begin{equation} \label{w-3}
-\frac{\omega'}{c}l' \;=\; \omega \gamma\!\left(\frac{v}{c^{2}} - \frac{l}{c}\right).
\end{equation}
We multiply the equation \eqref{w-3} through by $c$:
\begin{equation} 
-\omega' l' \;=\; \omega \gamma\!\left(\frac{v}{c} - l\right). \end{equation}

\begin{equation} \label{w-1}
\text{So:} \quad l' \;=\; -\,\frac{\omega}{\omega'} \gamma\!\left(\frac{v}{c} - l\right).    
\end{equation}
Now we use the transformation \eqref{w}, and rewrite it as: 

\begin{equation} \label{w-2}
\frac{\omega}{\omega'} \gamma \;=\; \frac{1}{(1 - \tfrac{l v}{c})}.
\end{equation}
Substituting equation \eqref{w-2} into equation \eqref{w-1}, we get \eqref{eqa}.

\noindent $\gamma$ cancels out in equation \eqref{eqa} or the dependence on $\gamma$ is absorbed into the frequency transformation $\omega'$ \eqref{w}. $l'$ is a direction cosine.}

\begin{equation} \label{eqa}
\boxed{l' \;=\; \frac{l - \tfrac{v}{c}}{1 - \tfrac{l v}{c}}.}
\end{equation}
Then we calculate the coefficient of $\eta$ and $\zeta$:%
\footnote{For $\eta$:
\begin{equation}
-\frac{\omega'}{c}m' = -\frac{\omega}{c}m
\;\;\Rightarrow\;\;
m' = \frac{\omega}{\omega'}\,m.    
\end{equation}
Substituting equation \eqref{w-2}, we get equation \eqref{b'}.
And for $\zeta$:

\begin{equation}
-\frac{\omega'}{c}n' = -\frac{\omega}{c}n
\;\;\Rightarrow\;\;
n' = \frac{\omega}{\omega'}\,n.    
\end{equation}
Substituting equation \eqref{w-2}, we get equation \eqref{c'}.}

\begin{equation} \label{b'}
\boxed{m' = \frac{m}{\gamma(1 - \tfrac{l v}{c})}.}    
\end{equation}

\begin{equation} \label{c'}
\boxed{n' = \frac{n}{\gamma(1 - \tfrac{l v}{c})}.}    
\end{equation}

Equating like coefficients of $\tau,\xi,\eta,\zeta$ between \eqref{Phi} and \eqref{Ph-k} yields: from the $\tau$--term, the frequency transformation \eqref{w}; from the $\xi$--term, the longitudinal direction--cosine transformation \eqref{eqa}; and from the $\eta,\zeta$--terms, the transverse direction--cosine transformations \eqref{b'}--\eqref{c'}. 

\item From the transformation of the angular frequency $\omega'$ \eqref{w}, Einstein derives the general Doppler formula \cite{Einstein05}:%
\footnote{In the rest frame $K$, the plane wave is written with direction cosines:
\begin{equation} \label{acphi}
l = \cos{\varphi}, \qquad  m =\cos{\theta}, \qquad n =\cos{\psi}.   
\end{equation}
Since Einstein chose the $x=\xi$-axis (the direction of relative motion between the frames), he singled out:

\begin{equation} \label{acos}
l=\cos\varphi.    
\end{equation}
We substitute \eqref{acos} into $\omega'$ \eqref{w}:

\begin{equation} \label{w1}
\omega' = \omega \gamma\!\left(1 - \frac{v}{c} \, \cos\varphi \,\right).  
\end{equation}
Dividing both sides by $2\pi$, yields the frequency $\nu'$ \eqref{Dop}.}

\begin{equation} \label{Dop}
\textit{Relativistic Doppler principle:} \qquad \nu' \;=\; \nu\,\frac{1-\tfrac{v}{c}\cos\varphi}{\sqrt{\,1-\tfrac{v^2}{c^{2}}\,}}.    
\end{equation}

Next, from the transformed direction cosine $l$ of the wave normal Einstein extracts the Aberration law \cite{Einstein05}:%
\footnote{Substituting equation \eqref{acos} into equation \eqref{eqa} transforms as:
\begin{equation} \label{Ab-1}
l' \;=\; \frac{\cos\varphi - \tfrac{v}{c}}{1 - \tfrac{v}{c}\cos\varphi}. 
\end{equation}
Since:  
\begin{equation} \label{acos'}
l'\;=\; \cos{\varphi}'
\end{equation}
we get the relativistic aberration formula \eqref{Ab}.}

\begin{equation} \label{Ab}
\textit{Relativistic Aberration law:} \qquad \cos\varphi' \;=\; \frac{\cos\varphi - \tfrac{v}{c}}{1 - \tfrac{v}{c}\cos\varphi}. 
\end{equation}

Now we interpret the symbols as observers would. In $K$ the wave normal makes an angle $\varphi$ with the $+x$--axis, so $l=\cos\varphi$. The observer at rest in $k$ moves through $K$ with velocity $v$ along $+x$. The measured angle in $k$ is $\varphi'$ with $l'=\cos\varphi'$ [equation \eqref{acos'}]. Inserting $l=\cos\varphi$ [equation \eqref{acos}] into \eqref{w} gives the general relativistic Doppler formula \eqref{Dop}. From \eqref{eqa} with equation \eqref{acos}, we obtain the aberration law \eqref{Ab}. 

\item If $\varphi=\tfrac{\pi}{2}$ (the source is $90^\circ$ from our velocity direction), then $\cos\varphi=0$, and equation \eqref{Ab} reduces to \cite{Einstein05}:%
\footnote{By the aberration equation \eqref{Ab}, we get:
\begin{equation}
\text{Equation \eqref{Ab}} \quad\stackrel{\cos\varphi=0}{=}\quad -\,\frac{v}{c}\,\neq\,0.    
\end{equation}.}

\begin{equation} \label{cos0}
\cos\varphi' \;=\; - \frac{v}{c}.    
\end{equation}
This means the light ray has an $x$-component opposite the observer’s motion. In other words, the ray is tilted by an angle $\approx \tfrac{v}{c}$ (in radians) from the transverse direction.%
\footnote{We can recover Bradley’s classical aberration result, if we start from Einstein’s aberration law \eqref{Ab}, expand to first order in $\tfrac{v}{c}$ (for $\tfrac{v}{c} \ll 1$), and get that the apparent direction is shifted toward the direction of motion by an angle: $\alpha\approx\tfrac{v}{c}\sin \varphi$. 

\noindent For the special case: $\varphi = \tfrac{\pi}{2} \Rightarrow \sin \varphi=1$ (the star is $90^\circ$ from the observer's velocity direction), we consider light arriving transverse to the motion in the rest frame, and get: $\alpha=\tfrac{v}{c}$. That is the familiar Bradley umbrella--rain effect: tilt your umbrella by $\tfrac{v}{c}$ result. This equals the maximum annual aberration angle.}

\item Finally, Einstein reduces the general Doppler formula to the simple longitudinal form ($\varphi = 0$, i.e., the source and the observer on the same line as the motion, with the source receding) \cite{Einstein05}:%
\footnote{\label{recap} We take the longitudinal case $\varphi = 0$. Then:  

\begin{equation} \label{gamnu}
\nu' = \nu \,\frac{1 - \tfrac{v}{c}}{\sqrt{\,1 - \tfrac{v^2}{c^2}\,}} = \gamma \nu \, (1- \frac{v}{c})
= \nu \,\frac{1 - \tfrac{v}{c}}{\sqrt{(1 - \tfrac{v}{c})(1 + \tfrac{v}{c})}} 
= \nu \,\sqrt{\frac{(1 - \tfrac{v}{c})^2}{(1 - \tfrac{v}{c})(1 + \tfrac{v}{c})}}. 
\end{equation}

\begin{equation}
= \nu \,\sqrt{\frac{(1 - \tfrac{v}{c})(1 - \tfrac{v}{c})}{(1 - \tfrac{v}{c})(1 + \tfrac{v}{c})}}
= \nu \,\sqrt{\frac{1 - \tfrac{v}{c}}{1 + \tfrac{v}{c}}}.    
\end{equation}

If instead the source approaches ($\varphi = \pi$, so $\cos \varphi = -1$), we get the complementary form:

\begin{equation} \label{DopL-2}
\nu' = \nu \,\frac{1 + \tfrac{v}{c}}{\sqrt{\,1 - \tfrac{v^2}{c^2}\,}} 
     = \nu \,\sqrt{\frac{(1 + \tfrac{v}{c})(1 + \tfrac{v}{c})}{(1 - \tfrac{v}{c})(1 + \tfrac{v}{c})}} 
     = \nu \,\sqrt{\frac{1 + \tfrac{v}{c}}{1 - \tfrac{v}{c}}} \,.  \quad \textit{Doppler Principle (approaching)}  
\end{equation}}

\begin{equation} \label{DopL}
\textit{Doppler Principle (receding):} 
\qquad 
\nu' \;=\;\nu \,\sqrt{\frac{1 - \tfrac{v}{c}}{1 + \tfrac{v}{c}}},
\end{equation}
where Lorentz's formula is valid for $v \ll c$.%
\footnote{For small $\frac{v}{c} \ll 1$:
\begin{equation} \label{small}
\gamma = 1 + \tfrac{1}{2}\frac{v^2}{c^2} + O(\frac{v^4}{c^4}). 
\end{equation}
Plugging equation \eqref{small} into equation \eqref{gamnu}, yields:
\begin{equation}
\frac{\nu'}{\nu} = \gamma (1 \pm \frac{v}{c})
= \bigl[1 + \tfrac{1}{2}\frac{v^2}{c^2} + \tfrac{1}{2}\frac{v^3}{c^3}+ O(\frac{v^4}{c^4})\bigr](1 \pm \frac{v}{c}) = 1 \pm \frac{v}{c} + O(\frac{v^2}{c^2}). \\\ \text{So to first order:}        
\end{equation}
\begin{equation}
\nu'\approx \nu \left(1 \pm \frac{v}{c} \right), \\\ \text{which is Lorentz's classical Doppler formula \eqref{FT}.}   
\end{equation}}

\item \label{conclusion} Einstein concludes: “One sees that — in contrast to the usual conception — for $v=-c, \nu'=\infty$" \cite{Einstein05}.

In other words, “in contrast to the usual conception,” where one would expect linear behavior, instead, relativity forces these limiting divergences. 
\end{enumerate}

Two special cases follow from the Doppler general law \eqref{Dop}: 

\medskip

I) \emph{1905}: The pure longitudinal case $\varphi=0$ or $\pi$, giving the longitudinal Doppler formulas \eqref{DopL} [and its approaching variant \eqref{DopL-2}] \cite{Einstein05}.%
\footnote{Setting $l=\cos\varphi=\pm 1$ in the general Doppler principle \eqref{Dop} gives the purely longitudinal cases (see comment \ref{recap}): 1) \emph{Receding source:} $\varphi=0$ (wave normal along $+x$ in $K$), \eqref{DopL}. This is a redshift.
2) \emph{Approaching source:} $\varphi=\pi$ (wave normal along $-x$ in $K$), which is \eqref{DopL-2}. This is a blueshift.}

II) \emph{1907}: The pure transverse Doppler case/time–dilation effect, the case   $\varphi'=\pi/2$ in system $k$ (recall that the source is at rest in $K$ and the observer is at rest in $k$, the "moving system"). The general Doppler equation \eqref{Dop} reduces to:%
\footnote{We start with the general Doppler equation \eqref{Dop} and the aberration law \eqref{Ab}. We then impose the condition in the observer's frame $k$:  $\varphi'=\pi/2 $ so  $\cos\varphi'=0$.
Inserting this into the aberration equation \eqref{Ab}, we get:
\begin{equation}
0 \;=\; \frac{\cos\varphi - \frac{v}{c}}{1 - \frac{v}{c}\cos\varphi}   \quad\Rightarrow\quad 0 \cdot (1 - \frac{v}{c}\cos\varphi) = \cos\varphi - \frac{v}{c} \quad \Rightarrow
\quad \cos\varphi \;=\; \frac{v}{c}. 
\end{equation}
We substitute this into the general Doppler law \eqref{Dop}:
\begin{equation}
\nu' \;=\; \nu\,\frac{\,1 - \frac{v}{c}\cdot\frac{v}{c}\,}{\sqrt{1-\frac{v^2}{c^2}}}
\;=\; \nu\,\frac{\,1 - \frac{v^2}{c^2}\,}{\sqrt{1-\frac{v^2}{c^2}}}
\;=\; \nu\,\sqrt{1-\frac{v^2}{c^2}},  
\end{equation}
which is equation \eqref{TDI}.}

\begin{equation} \label{TDI}
\nu' \;=\; \nu \sqrt{1-\frac{v^{2}}{c^{2}}}.    
\end{equation}

In his 1907 review article, "On the Relativity Principle and the Conclusions Drawn from It," Einstein stated that, observed from the observer's reference system, the moving clock’s rate is smaller by the ratio $\,\sqrt{1-\tfrac{v^{2}}{c^{2}}}\,$, [equivalently its period is dilated by the factor $\gamma$ in \eqref{TD}] \cite{Einstein07}.%
\footnote{We regard the monochromatic source at rest in $K$ as a clock with proper frequency $\nu = 1/T$ and proper period $T$. 
From \eqref{TDI} we have $\dfrac{\nu}{\gamma}$.
Hence the observed \emph{period} in $k$ is \emph{increased} (dilated) by the factor $\gamma$:
\begin{equation}\label{TD}
T' \;=\; \frac{1}{\nu'} \;=\; \frac{1}{\nu}\,\frac{1}{\sqrt{1-\frac{v^{2}}{c^{2}}}}
\;=\; \gamma\,T \;=\; \frac{T}{\sqrt{1-\frac{v^{2}}{c^{2}}}}.
\end{equation}
Equivalently, the moving clock’s \emph{rate} is reduced by the factor $1/\gamma$, i.e., a clock moving uniformly with speed $v$ runs slow by the factor $1/\gamma=\sqrt{\,1-\tfrac{v^{2}}{c^{2}}\,}$ in its \emph{rate}:
\begin{equation}\label{rate}
\frac{\text{rate}_\text{moving}}{\text{rate}_\text{rest}}
\;=\; \frac{\nu'}{\nu}
\;=\; \sqrt{1-\frac{v^{2}}{c^{2}}},
\end{equation}
which is equation \eqref{TDI}.} 

\subsection{Lorentz's Doppler Darkness Paradox} \label{Dark}

I) \emph{Einstein's relativistic longitudinal Doppler law \eqref{DopL}}. The difference between Lorentz's "usual conception" \eqref{FT} (Einstein's remark \ref{conclusion}) and Einstein's relativistic longitudinal Doppler law \eqref{DopL} jumps out:  

The Doppler equations \eqref{Dop} and \eqref{DopL} are not just a technical correction of Lorentz's equation \eqref{FT}. The gap is conceptual, not merely technical. The relativistic Doppler and aberration equations \eqref{Ab} demonstrate that Lorentzian dynamics can at best sustain bounded Doppler behavior from equation \eqref{FT}. At the same time, only the kinematical approach grounded in the heuristic relativity principle and the Lorentz transformation \eqref{TI} entails unbounded behavior as $v \rightarrow \pm c$. 
    
    The classical Doppler formula [equation \eqref{FT}] is linear and modest in its predictions. When we contrast the relativistic longitudinal Doppler law \eqref{DopL} with the classical Doppler law \eqref{FT}, the difference becomes dramatic upon examining the limits:%
    \footnote{The Lorentzian-Galilean wave picture gives the Doppler equation:

\begin{equation} \label{nuL}
\nu' \;\simeq\; \nu \left( 1 + \frac{v\cos \varphi}{c} \right), 
\end{equation}
where $c$ is the propagation speed of light (in the Lorentzian-Galilean theory, it is the ether speed), and $v$ is the velocity of the source/observer.
If motion is along the line of sight, then $\varphi=0$, and we obtain Lorentz's Doppler equation.}  
If we let $v \rightarrow c$ (approaching) in "the usual conception", i.e., Lorentz's equation \eqref{FT}, the frequency doubles $\nu' \rightarrow 2 \nu$. 
However, if $v \rightarrow -c$ (receding at the speed of light), the frequency would vanish (see section \ref{Asym}):

\begin{equation} \label{zero}
\nu' = \nu \left( 1 + \frac{v}{c} \right) = \nu \left( 1 - \frac{c}{c} \right) = 0.
\end{equation}
From equation \eqref{zero}, in the limit $v \rightarrow -c$, the observed frequency tends to zero. The period goes to infinity, and the wave crests never catch up to the observer, so the oscillation rate vanishes. Practically, if $v=-c$, the observer would see no oscillation. In physical terms, the observer would not detect any light at all — complete redshift, which is effectively darkness. 
Conceptually, Lorentz still used the Galilean addition of velocities law. In that framework, there is no formal prohibition against writing $v=-c$. One can, at least algebraically, imagine an observer moving at the speed of light. Thus, according to Lorentz's formula \eqref{zero}, there appears to be an observer moving at the speed of light. 

However, Maxwell’s theory states that electromagnetic waves propagate with a speed $c$ in the ether. If an observer moves with $v=-c$ relative to the ether in the same direction as the wave, then relative to that observer, the light would stand still (standing wavefronts, frequency zero). That is impossible within Maxwell’s equations. So the situation was paradoxical even in the ether framework. At the age of 16, Einstein imagined moving alongside a light wave at $v=c$. In that case, we would see electric and magnetic fields oscillating sinusoidally in space, but constant in time. It is not darkness (no crests arrive); it is a contradiction. This is \emph{Einstein's chasing a light beam paradox at $v=c$}.

Since the Galilean kinematics did not forbid $v=-c$, the \emph{darkness} case was a fundamental problem in Lorentz’s framework. 
Lorentz's formula allows us to plug in a light-speed observer, and predicts a vanishing frequency as if massive bodies could travel at $c$. 

In section \S7, Einstein solves \emph{Lorentz's Doppler darkness paradox} as follows. Instead of capping at $2 \nu$ or going to exactly zero at $v \pm c$, he shows a smooth, asymptotic shift. Einstein's relativistic longitudinal Doppler law \eqref{DopL} shows that as $v \rightarrow c$, the observed frequency tends to zero, so the wave is redshifted away. Still, \emph{no observer with mass can ever reach $v=c$}, so these divergences are merely asymptotic. Conversely, as $v \rightarrow -c$, the observed frequency tends to infinity, so the wave is infinitely blueshifted, \emph{but never realized physically}. This is the essence of Einstein's comment \ref{conclusion}. The relativistic law exhibits radical divergences at the limiting velocity, which are direct consequences of the principle of relativity, Einstein's new kinematics, and the Lorentz transformations \eqref{TI}.

\medskip

II) \emph{Einstein's transverse relativistic Doppler law \eqref{TDI}}. In 1905, Einstein enforced reciprocity and symmetry, and then, in 1907, he explicitly pointed to the transverse Doppler effect \eqref{TDI} as a test of time dilation. 

That conceptual re-founding is an achievement, and not a tweak of Lorentz’s Doppler law \eqref{FT}. One cannot obtain the transverse Doppler shift \eqref{TDI} (or time dilation) while retaining Galilean velocity addition as a physical law. 
The pre-relativistic dynamic approach that retained \emph{Galilean} velocity addition yields only the
first-order (classical) Doppler law and no transverse effect. One obtains the transverse Doppler and the correct aberration law only by renouncing the Galilean addition law and treating the Lorentz transformation as the kinematics for clocks, rods, and waves.
Hence, Einstein did not just port a sound formula to light. His derivation of the relativistic Doppler effect was not a mere generalization of Lorentz's classical Doppler law \eqref{FT}. He rebuilt the kinematics from which he derived the relativistic Doppler and aberration laws, including the transverse Doppler effect (pure time dilation). 

\subsection{Doppler and Aberration à la Poincaré?} \label{RDAP}

Poincaré writes the Lorentz transformation, which gives us relations between coordinates $(x, t, y, z)$ and $(x', t', y', z')$ (where $c=1$) \cite{Poincare}:

\begin{equation} \label{eq97} 
x' = k \, (x + \varepsilon t), \quad t' = k \, (t + \varepsilon x), \quad y' = y, \quad z' = z, 
\end{equation} 
\begin{equation} 
\text{with:} \qquad k = \frac{1}{\sqrt{1 - \varepsilon^2}}, 
\end{equation}

\medskip

We begin with a monochromatic plane wave in $(x, y, z)$ of phase $\Phi$: 

\begin{equation} \label{phaseP}
\Phi=\omega t - n_x x - n_y y - n_z z.
\end{equation}
We denote by $\mathbf{n}=(n_x,n_y,n_z)$ the wave normal, normal to the wavefronts, in the direction of propagation, and write:
\begin{equation} \label{inv}
\omega t - n_x x - n_y y - n_z z
= \omega' t' - n'_x x' - n'_y y' - n'_z z'.    
\end{equation}
Now, \emph{that is retro-fitting Einstein’s conceptual step into Poincaré’s framework}. It is not something Poincaré himself could have justified in 1905. Here, we declare the phase to be invariant under Lorentz transformations \eqref{eq97}. In modern language, this means the phase is a relativistic scalar. This is not just a calculation move. Making the phase invariant was a conceptual step, not a purely technical one.
That is a very Einsteinian step because it assumes that all inertial frames are equivalent and that no preferred ether frame exists. The wave’s oscillations look the same to all observers [see Einstein's equation \eqref{Ph-k} and stage \ref{same} in his derivation]. Assuming the phase to be invariant \eqref{inv}, we can write down the Doppler and aberration formulas \eqref{DAP} in relativistic form by substituting Poincaré's Lorentz transformation formulas \eqref{eq97} into the phase \eqref{phaseP} and equating coefficients of $t,x,y,z$: terms in the equation derived by the Lorentz transformation \eqref{eq97} = terms in equation \eqref{inv}.%
\footnote{This yields the transformation of the wave normal and frequency:
\begin{equation}
\omega' = k\bigl(\omega - \varepsilon\,n_x\bigr), \qquad  n'_x = k\bigl(n_x - \varepsilon\,\omega\bigr), \qquad n'_y = n_y, \qquad n'_z = n_z.  
\end{equation}
from which the relativistic Doppler and stellar aberration are obtained:
\begin{equation} \label{DAP}
\ \nu' = k\,\nu\,(1-\varepsilon\cos\theta), \qquad (\nu=\omega/2\pi), \qquad \text{and:} \quad \cos\theta'=\frac{\cos\theta-\varepsilon}{1-\varepsilon\cos\theta}.
\end{equation}}

Poincaré \emph{would not} naturally make that assumption. Poincaré, in 1905, was still thinking within Lorentz’s electron theory. His Lorentz transformations \eqref{eq97} are a clever mathematical tool that demonstrates no experiment can reveal absolute motion relative to the ether. From that point of view, the phase of a light wave \eqref{phaseP} would be "attached" to the ether frame. Thus, for Poincaré, to assert that the phase is invariant would have been tantamount to denying the ether’s privileged status — something he never did. That is the reason why Poincaré had all the mathematics in his hands (Lorentz transformation, group property, etc.), but did not derive the relativistic Doppler or aberration formulas the way Einstein did.

\section{Fizeau and Fresnel}

\subsection{Fizeau's Experiment}

Suppose light propagates in water at speed $c/n$ (where $n$ is the refractive index of water and $c$ is the speed of light in vacuum) in the rest frame of the water. The water itself is flowing at speed $v$ relative to the lab frame (where the apparatus is at rest). Then, according to the Galilean addition law of velocities, an observer in the lab frame measures the light speed:

\begin{equation} \label{eq18}
u = \frac{c}{n} + v. 
\end{equation}

However, in 1851, Hippolyte Fizeau measured the speed of light in moving water and discovered that the velocity of light in water does not fully obey the Newtonian velocity addition formula \eqref{eq18} \cite{Fizeau1851}, \cite{Fizeau1859}. Instead, Fizeau's experiment confirmed Augustin-Jean Fresnel's formula \cite{Fresnel1818}:

\begin{equation} \label{eq21}
u = \frac{c}{n} + v (1 - \frac{1}{n^2}), 
\end{equation}
where the factor:

\begin{equation} \label{eq20}
1 - \frac{1}{n^2}, 
\end{equation}
is called the Fresnel dragging coefficient.

This result showed partial dragging of the light wave by the moving water, which classical mechanics did not predict. It indicated that light’s speed was not simply as expected from Newtonian mechanics, as given by equation \eqref{eq18}. Fizeau’s results supported the idea that light’s velocity is modified according to Fresnel’s formula. 

\subsection{Lorentz Derives Fresnel's Formula} \label{LF}

In 1895, Lorentz attempted to reconcile Fresnel’s dragging formula. Lorentz's response to Fizeau's water tube experiment was to think that light waves were partially dragged by the dielectric, but the ether itself remained immobile. 
Lorentz begins with the wave in the medium at rest (relative to the ether) \cite{Lorentz1895}:

\begin{equation} \label{eq:79}
\begin{aligned}    
&\Phi = A \cos \frac{2 \pi}{T} \left(t - \frac{b_x x + b_y y + b_z z}{W} + B \right). \\ 
&\boxed{\varphi= t - \frac{\mathbf{b}\!\cdot\!\mathbf{r}}{W} + B.} \quad \rightarrow  \quad \Phi = A \cos \frac{2 \pi}{T} \varphi.
\end{aligned}
\end{equation}
The medium is in uniform motion. 
After imparting to the medium a uniform velocity $\mathbf{p}$, Lorentz’s \emph{theorem of corresponding states} says that a solution of the same form exists in the moving medium when one replaces $t$ by the \emph{local time} \cite{Lorentz1895}:

\begin{equation} \label{eq19}
t' = t- \frac{p_x x + p_y y + p_z z}{V^2}. \qquad t'=t - \frac{\mathbf{p}\!\cdot\!\mathbf{r}}{V^2}.    
\end{equation}
Lorentz writes \cite{Lorentz1895}:
\begin{equation} \label{eq:79-1}
\Phi = A \cos \frac{2 \pi}{T} \left(t' - \frac{b_x x + b_y y + b_z z}{W} + B \right). \qquad
\boxed{\varphi= t' - \frac{\mathbf{b}\!\cdot\!\mathbf{r}}{W} + B.}    
\end{equation}
In equation \eqref{eq:79-1}, Lorentz introduces the local time $t'$ \eqref{eq19} to describe what happens in the moving system. He is basically re-expressing the phase as it would appear in coordinates adapted to the moving system. 
Thus, the phase \eqref{eq:79} acquires, to first order in $\frac{|\mathbf{p}|}{V}$, the correction \cite{Lorentz1895}:

\begin{equation}
\begin{aligned} \label{eq:80}  
&\Phi = A \cos \frac{2 \pi}{T} \left(t-\frac{p_x x + p_y y + p_z z}{V^2} - \frac{b_x x + b_y y + b_z z}{W} + B \right). \\
& \boxed{\varphi = t \;-\; \frac{\mathbf{p}\!\cdot\!\mathbf{r}}{V^{2}} \;-\; \frac{\mathbf{b}\!\cdot\!\mathbf{r}}{W} \;+\; B.}  
\end{aligned}
\end{equation}
This is the essential operative step. The term $\frac{p_x x + p_y y + p_z z}{V^2}$ comes from the local time \eqref{eq19}. It is an additional correction to the phase.

\noindent Then Lorentz says that new direction cosines $b'_x, b'_y, b'_z$ of the wave normal are proportional to:

\begin{equation}
\frac{b_x}{W} + \frac{p_x}{V^{2}}, \qquad
\frac{b_y}{W} + \frac{p_y}{V^{2}}, \qquad
\frac{b_z}{W} + \frac{p_z}{V^{2}}.    
\end{equation}
So that we get \cite{Lorentz1895}:

\begin{equation} \label{eq:81}
\begin{aligned}    
&\frac{b_x}{W} + \frac{p_x}{V^{2}} = \frac{b'_x}{W'}\,,
\quad
\frac{b_y}{W} + \frac{p_y}{V^{2}} = \frac{b'_y}{W'}\,,
\quad
\frac{b_z}{W} + \frac{p_z}{V^{2}} = \frac{b'_z}{W'}, \\ &\displaystyle \boxed{\frac{\mathbf{b}}{W} + \frac{\mathbf{p}}{V^{2}} = \frac{\mathbf{b}'}{W'}.}
\end{aligned}
\end{equation} 
Substituting the right-hand side of equation \eqref{eq:81} into the right-hand side of equation \eqref{eq:80}, Lorentz gets \cite{Lorentz1895}:

\begin{equation} \label{eq:PI}
\begin{aligned}   
&\Phi = A \cos \frac{2 \pi}{T} \left(t - \frac{b'_x x + b'_y y + b'_z z}{W'}   + B \right). \\
&\boxed{\varphi= t - \frac{\mathbf{b}'\!\cdot\!\mathbf{r}}{W'} + B\,.}
\end{aligned}
\end{equation} 
Lorentz comments that "one sees that $W'$ is the velocity with which waves of the \emph{relative} oscillation period $T$ propagate in the direction $(b'_x, b'_y, b'_z)$ in the moving body" \cite{Lorentz1895}. Lorentz's derivation \emph{demonstrates} that the phase in the moving system \eqref{eq:PI} has the same functional form as the rest-frame plane wave \eqref{eq:79}, in line with Lorentz's 1895 “corresponding states” program.

Now, Lorentz re-expresses \eqref{eq:80} as a plane wave with a shifted normal and speed. Then, he extracts the speed $W'$, the light speed \emph{relative to the moving medium}.
To obtain the speed relative to the ether $W''$, Lorentz added back the component of the medium's drift along the ray \cite{Lorentz1895}:%
\footnote{We start from the vector relation that comes from \eqref{eq:81} and take norms using $|\mathbf{b}|=|\mathbf{b}'|=1$ ($\mathbf{b}$ and $\mathbf{b}'$ are direction cosines, unit vectors), and then square:
\begin{equation}
\left\lVert \frac{\mathbf b}{W} + \frac{\mathbf p}{V^{2}} \right\rVert^{2} = \left\lVert \frac{\mathbf b'}{W'} \right\rVert^{2} = \frac{1}{W'^2},     
\end{equation}

Then, we expand the left-hand side ($p_n \equiv \mathbf p \cdot \mathbf b$) \cite{Lorentz1895}:
\begin{equation}
\left(\frac{\mathbf b}{W} + \frac{\mathbf p}{V^{2}} \right)\!\cdot\!
\left(\frac{\mathbf b}{W} + \frac{\mathbf p}{V^{2}} \right) = \frac{\mathbf b\!\cdot\!\mathbf b}{W^{2}}
+ \frac{2\,\mathbf b\!\cdot\!\mathbf p}{W V^{2}}
+ \frac{\mathbf p\!\cdot\!\mathbf p}{V^{4}} = \frac{1}{W^{2}} + \frac{2\,p_n}{W V^{2}} + \frac{|\mathbf p|^{2}}{V^{4}}. 
\end{equation}
We neglect terms of the second order $O(p^{2}/V^{2})$, which gives \cite{Lorentz1895}:
\begin{equation}\label{eq:82}
\frac{1}{W'^2}
= \frac{1}{W^{2}} + \frac{2\,p_n}{W V^{2}}
= \frac{1}{W^{2}}\!\left(1 + \frac{2\,p_n W}{V^{2}}\right).
\end{equation}
From \eqref{eq:82}, we get:
\begin{equation} \label{eq:85}
\frac{W'}{W} = \left(1 + \frac{2\,p_n W}{V^{2}}\right)^{-1/2}.    
\end{equation}
Let $\varepsilon = \dfrac{2\,p_n W}{V^{2}}$. Since $\mathbf p \ll V, \varepsilon \ll 1$, $\dfrac{1}{\sqrt{1+\varepsilon}} \approx 1 - \dfrac{\varepsilon}{2}$. Therefore, to first order, \eqref{eq:85} becomes \cite{Lorentz1895}:

\begin{equation}\label{eq:83}
W' \approx W\!\left(1 - \frac{1}{2}\,\frac{2\,p_n W}{V^{2}}\right)
= W - \frac{p_n W^{2}}{V^{2}}. 
\end{equation}
Using $W= \dfrac{V}{N}$ (i.e., $\dfrac{W^2}{V^2} = \dfrac{1}{N^2}$), we get: $W'= W - \dfrac{p_n}{N^{2}}$.
Now we add the medium’s drift along the ray of light (a first-order addition), equation \eqref{eq:84}}.

\begin{equation} \label{eq:84}
W'' = W' + p_n \;=\; \,W + \Bigl(1 - \dfrac{1}{N^{2}}\Bigr)p_n,  \qquad W= \dfrac{V}{N},   
\end{equation}
which is Fresnel's formula.

\subsection{Retrofitting Lorentz's Derivation} \label{retro}

By retrofitting Lorentz’s derivation, one can recover a first-order analogue of the velocity addition law. Starting from equation \eqref{eq57}, which arises under the assumption of an isotropic medium and the neglect of terms beyond the first order in $\frac{\mathbf{p}}{V}$, the wave speed in the moving medium may be written in the form:%
\footnote{It comes from the following path to Fresnel's formula (which Lorentz does not follow). Equation \eqref{eq:81}($\mathbf{b}'=\mathbf{b}$) gives:
\begin{equation} \label{eq19-1}
\frac{1}{W'} \;=\; \frac{1}{W} \;+\; \frac{p_n}{V^2}. \quad \text{Inverting to first order:} \quad
W' = \frac{1}{\dfrac{1}{W} + \dfrac{p_n}{V^2}} = \frac{1}{\dfrac{1}{W}\left(1 + \dfrac{W p_n}{V^2} \right)}.
\end{equation}
We use a first-order expansion $\dfrac{1}{1+x} = 1 \, - x \, + x^2\, - \ldots \simeq 1 - x$ and set $x =  \dfrac{W p_n}{V^2}$. For $|x| << 1$, hence: 

\begin{equation}
W'\simeq W \left( 1 - \frac{W p_n}{V^2}\right), \qquad \text{dropping:} \qquad O\left(\frac{W^2 p^2_n}{V^4}\right)  \qquad \text{terms.} 
\end{equation}

\begin{equation} \label{eq57}
W' \;=\; \frac{W}{1 + \dfrac{W p_n}{V^2}}
\;\simeq\;
W \left( 1 \;-\; \frac{W\,p_n}{V^2} \right)
\;=\;
W \;-\; \frac{W^2}{V^2}\,p_n .
\end{equation}}

\begin{equation} 
W' \;=\; \frac{W}{1 + \dfrac{W p_n}{V^2}}.
\end{equation}
We add the drift velocity of the medium along the ray \eqref{eq:84} $W'' = W' + p_n$. 
Dropping higher-order terms $O(\frac{p_n^2}{V^2})$ (which is Lorentz's first-order scheme), gives:%
\footnote{Now we put everything over the common denominator:

\begin{equation}
W''= \frac{W}{1+\frac{Wp_n}{V^2}} + p_n = \frac{W + p_n\left(1+\frac{Wp_n}{V^2}\right)}{\,1+\frac{Wp_n}{V^2}\,} =  \frac{W + p_n + \frac{W p_n^2}{V^2}}{\,1+\frac{W p_n}{V^2}\,}.
\end{equation}

\begin{equation} 
W''= \frac{W+p_n}{\,1+\frac{Wp_n}{V^2}\,} \;+\; \frac{\tfrac{W p_n^2}{V^2}}{\,1+\frac{Wp_n}{V^2}\,} = \frac{W+p_n}{\,1+\frac{Wp_n}{V^2}\,} \;+\; O\!\left(\frac{p_n^2}{V^2}\right).   
\end{equation}}

\begin{equation} \label{eq58}
W'' \simeq \frac{W+p_n}{\,1+\dfrac{Wp_n}{V^2}\,}.
\end{equation}

\subsection{Einstein Rederives the Addition Law} \label{1907} 

In his 1907 review article, Einstein revisited the derivation he presented in 1905. This time, he extended it to refractive media, reintroduced Fizeau’s experiment, and showed that the same phase-invariance method yields the \emph{exact} relativistic velocity-addition law for phase velocity. Einstein's 1907 methodology is identical to his 1905 procedure; the only difference is that Einstein now makes the velocity addition law explicit. However, when applied to media, the law reduces to Fresnel’s drag to first order.

Einstein began with the phase \eqref{eq:phase_Sprime} in the rest frame $S'$ and transformed it to the laboratory frame $S$ using the Lorentz transformation \eqref{eq:LT}, obtaining \eqref{eq:phase_S}. By the principle of relativity, the phase in $S$ must still take the plane–wave form \eqref{Rel}. Comparing \eqref{eq:phase_S} with \eqref{Rel}, Einstein derived the relations \eqref{eq:omega_u_relations}. Eliminating $\omega$ and $\omega'$ gives the exact relativistic addition law for phase velocity \eqref{eq:addition_u}. For $u' = c/n$, this reduces to \eqref{eqdisp}, which for $v \ll c$ yields Fresnel’s drag coefficient \eqref{eqFr}.

Let $S$ be the stationary (laboratory) system with coordinates ($x,y,z,t$) and $S'$ a system moving with constant velocity $v$ in the $+x$-direction relative to $S$,
with coordinates ($x',y',z',t'$). The liquid is at rest in $S'$ and is assumed homogeneous and isotropic in that system.

For a monochromatic plane wave in $S'$ propagating along $+x'$ with phase velocity $u'$, the phase may be written as \cite{Einstein07}:

\begin{equation}\label{eq:phase_Sprime}
\Phi \;=\; \omega'\!\left(t' \;-\; \frac{x'}{u'}\right).
\end{equation}
Between $S$ and $S'$ the Lorentz transformation holds:
\begin{equation}\label{eq:LT}
x'=\gamma(x-vt),\qquad
t'=\gamma\!\left(t-\frac{v}{c^{2}}x\right),\qquad
\gamma=\frac{1}{\sqrt{1-\tfrac{v^{2}}{c^{2}}}}.
\end{equation}
According to the principle of relativity, the phase in $S$ (the laboratory) is:

\begin{equation} \label{Rel}
\Phi \;=\; \omega\,t \;-\; \frac{\omega}{u}\,x.
\end{equation}
Substituting the Lorentz transformation \eqref{eq:LT} into the phase \eqref{eq:phase_Sprime},%
\footnote{collecting terms:

\begin{align}
t' - \frac{x'}{u'} 
&= \gamma\!\left(t - \frac{v}{c^{2}}x\right) \;-\; \frac{\gamma}{u'}\,(x-vt) \\[6pt]
&= \gamma\!\left[t - \frac{v}{c^{2}}x - \frac{x}{u'} + \frac{v}{u'}t\right] \\[6pt]
&= \gamma\!\left[\Bigl(1+\frac{v}{u'}\Bigr)t \;-\; \Bigl(\frac{1}{u'}+\frac{v}{c^{2}}\Bigr)x\right].
\end{align}}
and multiplying by $\omega'$ gives:

\begin{equation}\label{eq:phase_S}
\Phi
= \underbrace{\omega'\gamma\!\left(1+\frac{v}{u'}\right)}_{\displaystyle \omega}\,t
\;-\;
\underbrace{\omega'\gamma\!\left(\frac{1}{u'}+\frac{v}{c^{2}}\right)}_{\displaystyle \omega/u}\,x.
\end{equation}
We compare with the plane–wave form in the $S$ frame \eqref{Rel} to identify \cite{Einstein07}:
\begin{equation}\label{eq:omega_u_relations}
\omega=\omega'\gamma\!\left(1+\frac{v}{u'}\right),
\qquad
\frac{\omega}{u}\,=\, \omega'\gamma\!\left(\frac{1}{u'}+\frac{v}{c^{2}}\right).
\end{equation}
Now we divide them:

\begin{equation}
u\;=\;\frac{\omega}{\tfrac{\omega}{u}} \;=\; \frac{1+\dfrac{v}{u'}}{\dfrac{1}{u'}+\dfrac{v}{c^{2}}}\,.    
\end{equation}
Solving the second relation yields the (exact) velocity addition law for the phase velocity \cite{Einstein07}:

\begin{equation}\label{eq:addition_u}
u \;=\; \frac{u' + v}{\,1+\dfrac{u'v}{c^{2}}\,}.
\end{equation}
In his \emph{Jahrbuch der Radioaktivität und Elektronik} review paper, Einstein mentions Fizeau's experiment and hints at the derivation \cite{Einstein07}. He does not carry out the algebra himself in the review. Instead, he footnotes Max Laue’s paper \cite{Laue1907}, published earlier in 1907, which contains the explicit derivation. Since Einstein's review was intended as a summary rather than a research article, and because Laue had already published the derivation, Einstein simply indicated the result and referred readers to Laue for the full demonstration.
In 1907, Laue inserted $u'=\frac{c}{n}$ into the velocity-addition formula \eqref{eq:addition_u}, and obtained equation \cite{Laue1907}.%
\footnote{According to Einstein's methodology, we formulate the phase \eqref{eq:phase_Sprime} in the rest frame of the medium as:

\begin{equation}
\Phi \;=\; \omega'\!\left(t' \;-\; \frac{n}{c}\, x'\right) \;=\; \omega' \!\left(t' \;-\; \frac{x'}{u'}\right), \qquad \text{where} \quad u'=c/n,
\end{equation}
and then if the medium in $S'$ is effectively nondispersive with refractive index $n$ (so $u'=\frac{c}{n}$), equation \eqref{eq:addition_u} gives equation \eqref{eqdisp}.}

\begin{equation}\label{eqdisp}
w \;=\; \frac{\tfrac{c}{n}+v}{\,1+\dfrac{v}{nc}\,}
\;=\; \frac{c+nv}{\,n+\tfrac{v}{c}\,}.
\end{equation}
For $v\ll c \, $, he used the first-order Taylor expansion of $(1+x)^{-1}$, and got \cite{Laue1907}:

\begin{equation}\label{eqFr}
w \;\approx\; \frac{c}{n} \;+\; v\!\left(1-\frac{1}{n^{2}}\right)\;+\; O\!\left(\frac{v^2}{c^2}\right),
\end{equation}
Fresnel’s drag to first order (for $n=1$, $u=c$).

\section{Abolishing the Ether Kinematics} 

Einstein acknowledged that he had read Lorentz’s \emph{Versuch} \cite{Lorentz1895} well before 1905. By 1902, he had carefully studied the 1895 treatise, \emph{CPAE1}, Doc. 130. In that work, Lorentz had already derived the classical Doppler effect (discussed in section \ref{CD}) and invoked local time to derive Fresnel’s drag coefficient from Maxwell’s equations (discussed in section \ref{LF}). In his 1907 \emph{Jahrbuch} review, Einstein explicitly pointed to Lorentz’s 1895 \emph{Versuch} as a key precursor to relativity, emphasizing that although it still presupposed an immobile ether, the theory had gained empirical support—not least by correctly yielding Fresnel’s dragging coefficient in agreement with Fizeau’s experiment \cite{Einstein07}.
As my reconstructions of Lorentz’s 1895 and Einstein’s 1905-1907 derivations show, the "family resemblance" between them is genuine. I will summarize the main points here.

Lorentz’s 1895 optics of moving bodies was valid only to first order in $\tfrac{v}{c}$, sufficient to explain the classical Doppler effect and Fresnel’s drag \cite{Lorentz1895}. Einstein, by contrast, began from the phase of the wave in his kinematical treatment of the optics of moving bodies. This led directly to the relativistic Doppler and aberration laws and to the exact velocity addition formula \cite{Einstein05, Einstein07}, with Fresnel’s drag emerging as the $v \ll c$ limit \cite{Laue1907}. 

In section \ref{retro}, I show that by retrofitting Lorentz’s derivation, one can recover a first-order analogue of the velocity addition law. The result \eqref{eq58} resembles the relativistic law of addition of velocities \eqref{eq:addition_u}. Yet the resemblance is purely formal. It holds only to first order in $p_n/c$, since the starting point, equation \eqref{eq57}, was itself already a first-order approximation.
The similarity to Einstein’s formula is therefore accidental. In Einstein’s theory, the velocity addition law follows rigorously and exactly from the Lorentz transformations, valid to all orders in $\frac{v}{c}$. The relativistic law of velocity addition emerges immediately and without approximation. What looks structurally similar is thus no more than a first-order mimicry of relativity, whereas Einstein’s result is exact and grounded in a new heuristic.

The mathematical technique (starting with a rest–frame plane wave, mapping to a moving frame, and comparing coefficients) and the local–time structure clearly flow from Lorentz \cite{Lorentz1895} to Einstein \cite{Einstein05, Einstein07}.
But Einstein’s methodology could hardly be more different. Einstein converts Lorentz’s constructive recipe into a principle theory with phase invariance, Lorentz invariance, and no ether. The same Fresnel number emerges, but now as kinematics, not as a property of ether–matter interaction:

Lorentz's derivation is valid only to first order and uses the device of local time \eqref{eq19}.     Einstein's derivation is exact, employing the full Lorentz transformation \eqref{eq:LT} (and only then reduced to first order by recovering Fresnel's drag). The technical template — starting from a plane wave in the comoving frame, transforming it, and then reading off the frequency and wave number — is common to both. Einstein's equations \eqref{eq:phase_Sprime}, \eqref{eq:LT}, \eqref{eq:phase_S}, and \eqref{eq:omega_u_relations} implement this exactly in the collinear case. Lorentz's equations \eqref{eq:79}, \eqref{eq19}, \eqref{eq:81}, and \eqref{eq:PI} are the earlier version, worked out vectorially but truncated at first order. Lorentz \emph{derives} from the corrected phase the identity \eqref{eq:81} as an algebraic device. Einstein’s relations \eqref{eq:omega_u_relations} express the same content in the collinear case, but they fall out immediately by reading off the coefficients in \eqref{eq:phase_S} and matching to equation \eqref{Rel}.
    
Lorentz shows \emph{constructively} that the phase in the moving system can be recast into the same form as in the rest system in the ether, by redefining direction cosines $\mathbf{b}$ and speed $W$ [see equations \eqref{eq:81} $\to$ \eqref{eq:PI}]. In other words, Lorentz demonstrates phase invariance \eqref{eq:81} to first order via local time \eqref{eq19}. Equation \eqref{eq:81} is thus the vector, first-order ancestor of the exact coefficient relations \eqref{eq:omega_u_relations}. This is a \emph{derived}, ether-based, first–order "same form."   By contrast, Einstein \emph{requires} from the outset, by the principle of relativity, that the wave in any inertial frame has the plane–wave form \eqref{Rel}.     The equality of form is not demonstrated but \emph{postulated}, and the Lorentz transformations \eqref{eq:LT} then guarantee it exactly. 

This methodological contrast recalls Paul Ehrenfest’s remark after reading Einstein’s 1905 analysis of the electron mass. Ehrenfest interprets Einstein’s relativity as essentially Lorentz's theory reformulated, since the results coincide \cite{Ehrn}. In his reply to Ehrenfest, Einstein acknowledged the \emph{identity of results}, yet denied an \emph{identity of theories}. His relativity principle, he emphasized, was not a deductive "closed system" but a heuristic principle that guides connections across domains \cite{Einstein07}. Thus, Einstein’s contribution lies not in producing the same mathematical forms as Lorentz, but in the shift of conceptual foundation — from derivation within an ether theory to the postulation of a principle that guides and demands formal identity across inertial frames.

Einstein’s move was \emph{to change the kinematics}. He abolished the ether kinematics. He assumed from the outset the principle of relativity and treated the Lorentz transformations as exact, kinematical relations between inertial frames. For him, the phase invariance of a plane wave was not a device to salvage Maxwell’s equations, but a direct consequence of the principle of relativity. From this purely kinematical basis, he derived the exact relativistic velocity addition law, along with the relativistic Doppler effect and aberration.

According to Michel Janssen, Einstein’s 1905 innovation was not new mathematics but a new interpretation of Lorentz’s 1895 \emph{Versuch}. Where Lorentz regarded the invariance of Maxwell’s equations as a property tied to an immobile ether, Einstein elevated this invariance into a universal feature of nature grounded in the relativity principle and the constancy of light speed. In doing so, he supplied a common-cause explanation. The Lorentz invariance of both fields and matter was no longer an unexplained coincidence but a necessary consequence of the underlying spacetime structure: "...the new interpretation traced to a common cause what in the old interpretation were
unexplained coincidences" \cite{Janssen}.

When deriving the classical Doppler effect (see section \ref{CD}), Lorentz retained the Galilean velocity addition rule as the kinematics (so that an observer moving with velocity $v$ relative to the ether would “see” light at $c-v$). His substitution \eqref{sigma} is nothing but a Galilean change of variables, assuming $x' = x - vt$ and $t'=t$. This is the kinematical step that embodies Galilean addition. 
By contrast, in the same 1895 treatise, for electrodynamics in moving media, Lorentz introduced the local time to preserve the form of Maxwell’s equations (see section \ref{LF}) \cite{Lorentz1895}. The result was an uneasy hybrid. On the one hand, electrodynamics remained formally invariant by virtue of local time; on the other, the underlying kinematics still presupposed Galilean velocity addition, so that in principle one could “ride along” with a light wave and see it stand still if $v=c$. 

Einstein’s step in 1905 was far more radical. He rejected this hybrid and unified the optics and electrodynamics of moving bodies under a single kinematics. By deriving the Lorentz transformations from the relativity principle and the constancy of $c$, he obtained both the relativistic Doppler effect \cite{Einstein05} and the velocity addition law \cite{Einstein07} (and thus Fresnel’s drag \cite{Laue1907}) within one kinematical framework. In Lorentz’s work, the two phenomena required different prescriptions (Galilean kinematics for Doppler, local time for media) \cite{Lorentz1895}. Still, in Einstein’s relativity, they emerge as two aspects of the same relativistic kinematics. 

While there is a genuine influence of Lorentz on Einstein’s reasoning, the conceptual distance is profound. It is precisely the gap Einstein later described as that between a \emph{constructive theory} (Lorentz’s mathematical maneuvers on an ether background) and a \emph{principle theory} (special relativity, founded on general principles from which the formalism follows). 
Only then does the paradox vanish. We can never transform into a frame where a propagating light wave stands still, because $c$ is invariant and no inertial frame reaches $v=c$.

Poincaré went further than Lorentz by recognizing the group properties of the transformations and emphasizing the principle of relativity. Yet he too remained within the conceptual confines of the ether, with all the limitations this entailed. The absence of the relativistic Doppler and aberration laws in Poincaré’s electron dynamics cannot be explained simply by noting that he treated forces, stresses, and Lorentz's electron model, while Einstein focused on waves. The deeper reason is ontological. Like Lorentz, Poincaré retained the ether—conceived as a real, if undetectable, medium—as the ultimate backdrop of electrodynamics. This commitment barred him from taking the decisive step that Einstein did in 1905. 

It is therefore mistaken to suppose that both men possessed the same formal machinery and that Einstein merely applied it to different problems, such as Doppler or aberration. As I demonstrate in Section \ref{RDAP}, even with Poincaré’s full Lorentz transformation \eqref{eq97} \cite{Poincare}, those relativistic laws could not have been followed. Einstein’s achievement was not one of topic choice but of conceptual revolution. The relativistic Doppler and aberration formulas, which arise naturally in his 1905 paper, testify to something more profound: a wholesale rethinking of the framework itself. Where Poincaré extended Lorentz’s dynamics, Einstein established a new kinematics—and with it, a new conception of Newtonian classical mechanics.

Thus, the relativistic Doppler and aberration laws do not mark a different application of the same theory — they mark the point at which Einstein’s theory departs from Lorentz's and Poincaré’s theories altogether.

\section*{Acknowledgment}

I want to thank the late Prof. John Stachel for spending numerous hours with me discussing Einstein's relativity at the Center for Einstein Studies at Boston University.

\end{document}